\documentclass[aps,prx,onecolumn,noshowpacs,amsfonts,amssymb,amsmath,longbibliography,superscriptaddress,nofootinbib,notitlepage]{revtex4-2}

\usepackage{bm}
\usepackage{graphicx}
\usepackage[caption=false]{subfig}
\usepackage[utf8]{inputenc}
\usepackage{amsmath}
\usepackage[pdftex, colorlinks=true, allcolors=monbleu]{hyperref}
\usepackage{color}
\definecolor{monbleu}{RGB}{76, 114, 176}
\usepackage{longtable}
\usepackage{mathtools}
\def\multiset#1#2{\ensuremath{\left(\kern-.3em\left(\genfrac{}{}{0pt}{}{#1}{#2}\right)\kern-.3em\right)}}

\graphicspath{{./figs/}} 
\begin{document}
%
%
%
%
\title{Network compression with configuration models and the minimum description length}
\author{Laurent \surname{H\'ebert-Dufresne}}
\affiliation{Vermont Complex Systems Center, University of Vermont, Burlington VT}
\affiliation{Department of Computer Science, University of Vermont, Burlington VT}
\author{Jean-Gabriel \surname{Young}}
\affiliation{Vermont Complex Systems Center, University of Vermont, Burlington VT}
\affiliation{Department of Computer Science, University of Vermont, Burlington VT}
\affiliation{Department of Mathematics \& Statistics, University of Vermont, Burlington VT}
\affiliation{D\'epartement de physique, de g\'enie physique et d'optique, Universit\'e Laval, Qu\'ebec (Qu\'ebec), Canada G1V 0A6}
\author{Alexander Daniels}
\affiliation{Vermont Complex Systems Center, University of Vermont, Burlington VT}
\author{Alec Kirkley}
\affiliation{Institute of Data Science, University of Hong Kong, Hong Kong}
\affiliation{Department of Urban Planning and Design, University of Hong Kong, Hong Kong}
\affiliation{Urban Systems Institute, University of Hong Kong, Hong Kong}
\author{Antoine \surname{Allard}}
\affiliation{D\'epartement de physique, de g\'enie physique et d'optique, Universit\'e Laval, Qu\'ebec (Qu\'ebec), Canada G1V 0A6}
\affiliation{Centre interdisciplinaire en mod\'elisation math\'ematique, Universit\'e Laval, Qu\'ebec (Qu\'ebec), Canada G1V 0A6}
\affiliation{Vermont Complex Systems Center, University of Vermont, Burlington VT}
\date{\today}
%
\pacs{}
\begin{abstract}
Random network models, constrained to reproduce specific statistical features, are often used to represent and analyze network data and their mathematical descriptions. Chief among them, the configuration model constrains random networks by their degree distribution and is foundational to many areas of network science. However, configuration models and their variants are often selected based on intuition or mathematical and computational simplicity rather than on statistical evidence. To evaluate the quality of a network representation, we need to consider both the amount of information required to specify a random network model and the probability of recovering the original data when using the model as a generative process. To this end, we calculate the approximate size of network ensembles generated by the popular configuration model and its generalizations, including versions accounting for degree correlations and centrality layers. We then apply the minimum description length principle as a model selection criterion over the resulting nested family of configuration models. Using a dataset of over 100 networks from various domains, we find that the classic Configuration Model is generally preferred on networks with an average degree above ten, while a Layered Configuration Model constrained by a centrality metric offers the most compact representation of the majority of sparse networks.
\end{abstract}

\maketitle
%
%
%
%
%
\section{Introduction}
%
When we use random networks as models of empirical network data, we typically expect faithfulness to increase with model complexity. The idea is to include limited network properties that we believe are important while leaving everything else up to randomness. Hopefully, the more information we include, the more our random networks will look like the original network data, and our mathematical descriptions of the network will become more accurate.
For instance, consider the Erd\H{o}s-R\'enyi (ER) random graph model, the uniform distribution over all networks with a fixed number of nodes $N$ and edges $E$ \cite{erdos1960evolution}, which is perhaps the simplest of sketches one can make of a network.
A random realization of this model is unlikely to resemble an observed complex network, but the ER model will nonetheless reproduce its density if $N$ and $E$ are set to the observed values.
More complex models will capture more intricate patterns, such as degree heterogeneity or degree correlations.

\emph{Data compression} allows us to measure the effectiveness of any random network model by considering the cost of describing the model and the cost of describing an observed network given the information specified by the model. More detailed models (i.e., ones that impose more constraints on the network structure) are more costly to specify but provide a more precise summary of an observed network. 
For example, compressing network data to only two numbers, $N$ and $E$, provides a very succinct summary of the network, but given this information, we only know that the original network data is one of $\Omega = \binom{N(N-1)/2}{E} \sim N^N$ possible networks (scaling obtained by assuming the sparse limit where $E \sim N$). Any network sampled or generated from this ensemble is unlikely to be similar to the original network in any significant way. Hence, while this particular model is a useful mathematical simplification---we can compute the behavior of many network quantities mathematically in the ER model---it will not reproduce many important distinguishing features of the network it was meant to model.

\begin{figure*}
    \centering
    \includegraphics[width=0.8\linewidth, angle=0]{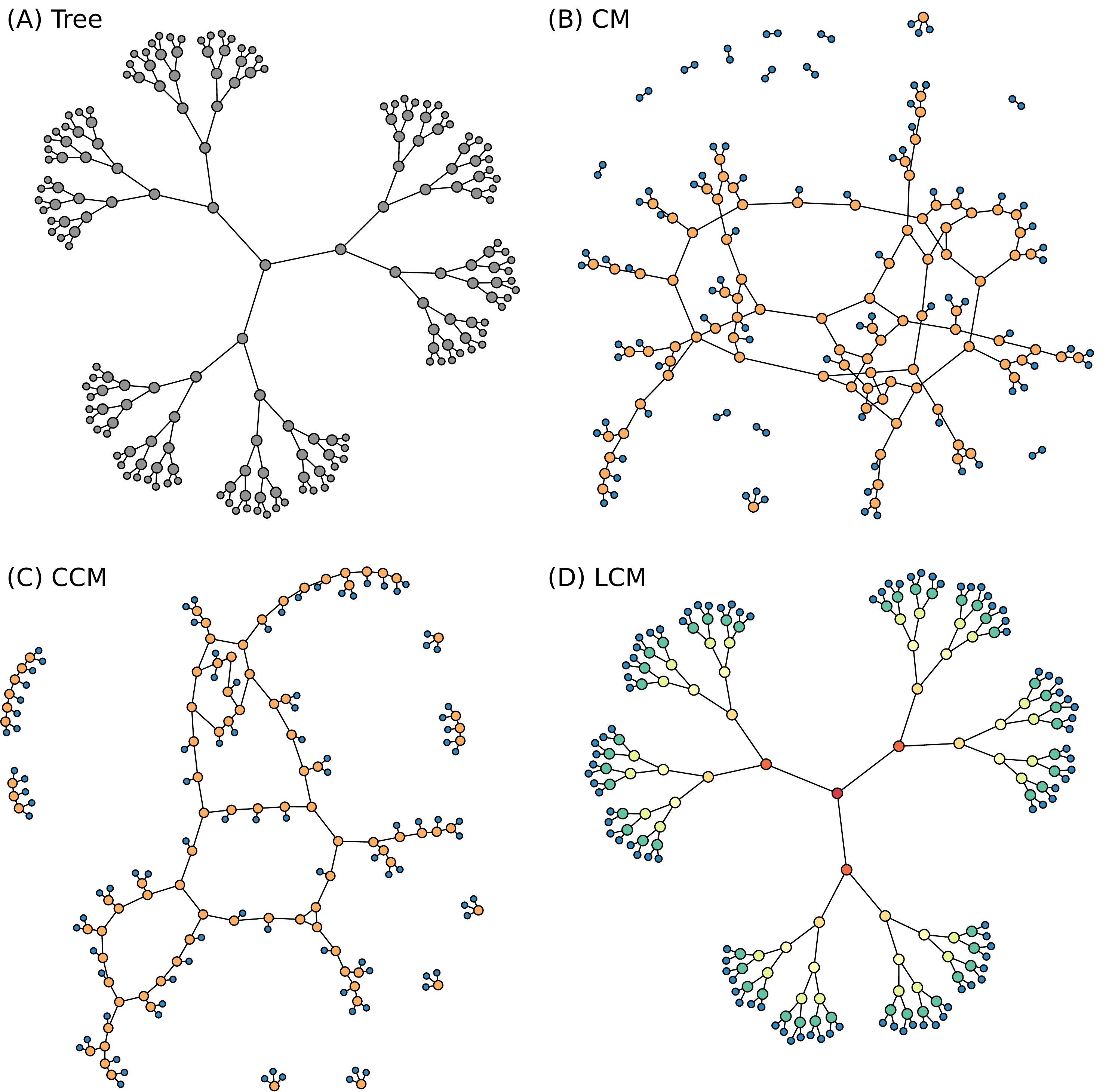}
    \caption{\textbf{Modeling a perfect tree.} (A) The original dataset, a Cayley tree. We show random realizations of two models that do not preserve the structure of the tree: (B) The Configuration Model (CM \cite{fosdick2018configuring}) preserving only the degree distribution (represented with different node colors) and (C) the Correlated Configuration Model (CCM \cite{vazquez2003resilience}) preserving joint degree-degree correlations also using two node types plus a 2-by-2 edge matrix specifying frequency of connections between them. (D) The original network is preserved by the Layered Configuration Model (LCM \cite{hebert2016multi}) using a node type for each of the 7 layers---the model can only generate networks with the same structure as the original tree up to relabeling. The Layered Correlated Configuration Model (LCCM \cite{allard2018percolation}), which is not shown, would be indistinguishable from the LCM shown in panel (D) as there are no degree-degree correlations not captured by the layer structure.}
    \label{fig:trees}
\end{figure*}

In this paper, we investigate the usefulness of new developments in Configuration Models~ \cite{newman2001random, fosdick2018configuring} in representing and compressing network data. 
These models incorporate simple observable structural features of networks. In its classic form, the Configuration Model (CM) summarizes a network with its degree sequence $N_k$: the number of nodes with $k$ edges. The ensemble of networks considered by the CM then contains all possible configurations that respect the original degree sequence. More complex models can be constructed in an analogous manner to add structure to the output, for example, degree-degree correlations~\cite{vazquez2003resilience} and centrality~\cite{hebert2016multi, allard2018percolation}; see Fig.~\ref{fig:trees} for some examples of models that will be investigated here. 

\begin{figure}
    \centering
    \includegraphics[width=\linewidth]{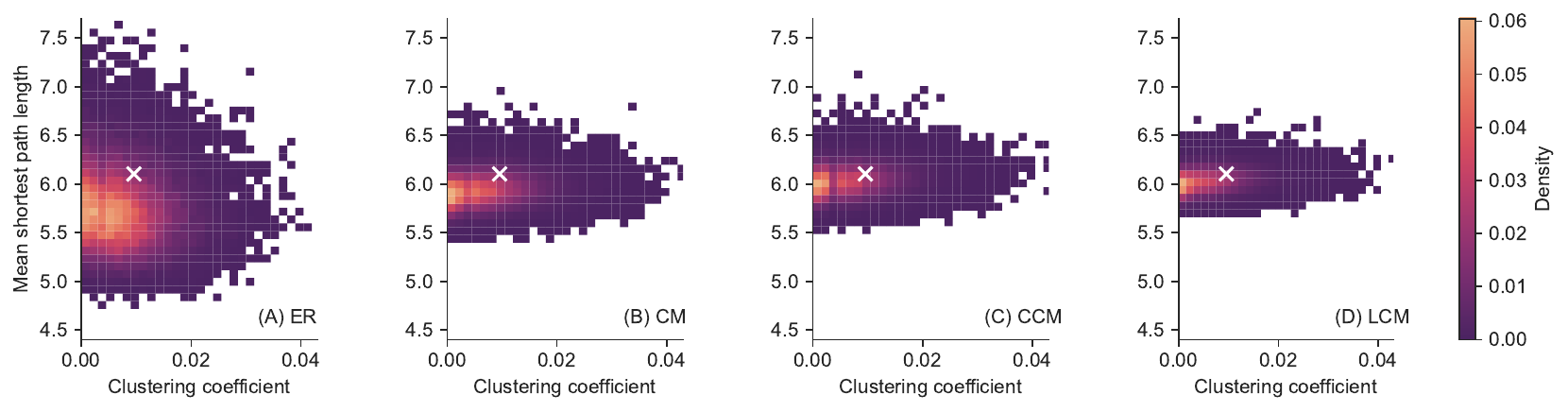}
    \caption{\textbf{Projection of random network ensembles.} We generated an ensemble of Erd\H{o}s-R\'enyi (ER) random graphs with $N=250$ nodes and $E=311$ edges (on average). We then plotted the density of graphs in the space defined by the clustering coefficient $C$ and the mean shortest path length $\ell$ of each graph \cite{Watts1998}. We then picked one unique graph at random (shown with a cross), and generated graphs from the corresponding CM, CCM and LCM ensembles, to see how the area covered by the ensembles in $(C,\ell)$-space would shrink with additional constraints. All ensembles are visualized with $1.4 \times 10^6$ random draws from each ensemble discretized into the bins indicated by grid cells in the figures.}
    \label{fig:space}
\end{figure}

The intuition motivating our analysis is that a smaller random network ensemble means a better representation of a network data set. Each network model defines a subset of the entire space of all possible network structures. A small ensemble size for the model means a smaller subset of that space and, therefore, smaller distances between any given random networks from the ensemble and the original network data. We visualize this intuition using a simple experiment in Fig.~\ref{fig:space}. We find without surprise that imposing different structural constraints on random network models changes the size and ``shape'' of the resulting network ensemble. What is less obvious, and therefore requires careful calculations, is the tradeoff between these models' accuracy and complexity. For example, in our simple experiment, the Layered Configuration Model (LCM) produced the ensemble with the least dispersal while requiring less information than the Correlated Configuration Model (CCM) but much more information than the CM. This paper formalizes this question within the nested family of Configuration Models: Which of these descriptive random network models best represents the structure of real complex networks?

To evaluate the quality of a representation more formally, we will use the minimum description length principle (DL)~\cite{grunwald2007minimum,mackay2003information,peixoto2013parsimonious}. This approach quantifies the level of compression a model provides by accounting for the amount of information (calculated in bits) needed to parametrize the model and the quality of compression the models provide once parametrized. In this framework, models that provide the most compression are preferred since good compression can only be achieved by identifying prominent patterns in the data and modeling these patterns explicitly. To prevent overly complex models from being preferred, the DL framework also considers the informational costs of specifying the model itself, meaning that, in practice, simple models that provide strong constraints on the structure are preferred---a formalization of Occam's razor. 
Description length methods have been used to provide compression of network data in a range of inference settings by exploiting structural regularities in network communities \cite{peixoto2019bayesian}, node features \cite{hric2016network,kirkley2022spatial,kirkley2024identifying}, and frequent interactions across network layers or time \cite{peixoto2017modelling,kirkley2023compressing,kirkley2024dynamic}.

Here, we derive description lengths for the Configuration models shown in Fig.~\ref{fig:trees}, and evaluate the compression obtained by these models on more than a hundred complex networks from different scientific domains, including, but not limited to, social networks (where nodes are often people and links are some interaction), biological networks (e.g., protein-protein interaction networks and phylogenetic trees), connectomes (e.g., axonal connections between brain regions), food webs, infrastructure networks (e.g., power grids) and transportation networks (e.g., flights between airports).
Our results show that complex configuration models can offer surprisingly strong compression, especially when networks are sparse.

%
%
%
%
%
\section{Methods}
%
\subsection{Nested family of Configuration Models}

The models considered in this work are the classic Configuration Model (CM \cite{newman2001random, fosdick2018configuring}), the Correlated Configuration Model (CCM \cite{vazquez2003resilience}), the Layered Configuration Model (LCM, previously introduced as the Onion Network Ensemble \cite{hebert2016multi}) and the Layered Correlated Configuration Model (LCCM \cite{allard2018percolation}). Examples of typical random instances of these models are shown in Fig.~\ref{fig:trees}. 

All of these models are special cases of a general random network model where (1) nodes are assigned one of many possible node types representing their local connectivity patterns, (2) edges follow an edge matrix specifying the number of edges between node types, and (3) additional connection rules can also be enforced by assigning types to stubs~\cite{allard2015general}. 

First, the types of nodes simply represent degree classes in the CM and the CCM, and they represent more complicated joint pairs of degree and centrality layer classes in the LCM and LCCM. These centrality layers are based on the Onion Decomposition \cite{hebert2016multi}, which is identical to the $k$-core decomposition but tracks the order by which the network is reduced when peeling nodes under a certain degree. We describe this idea further in the next sections, but importantly, this centrality structure fully specifies the structure of perfect trees and is therefore used to provide a treelike approximation of other networks \cite{allard2018percolation}.

Second, the CCM, LCM, and LCCM are also specified by an edge matrix specifying connections between types: degree-degree pairs in the CCM, layer-layer pairs in the LCM, and complex pairings of degree-layer pairs in the LCCM. These edge matrices help us specify correlations in otherwise random connections. This can help us reproduce, for example, the fact that social and technological networks both have heterogeneous degree distributions, but degree-degree correlations are expected to be positive in the former and negative in the latter \cite{newman2002assortative}. Without these correlations, data representing social or technological networks might look extremely different from networks generated by the simple Configuration Model. Hence, correlations can be critical to achieving a good model and an effective compression.

Third, the LCM and LCCM are also specified by a complex set of connection rules to preserve the centrality layer of nodes, whereas preserving degrees is relatively trivial in the CM and CCM. Within this general framework, we can calculate the number of possible networks in a given ensemble by thinking of stubs as tokens and estimating the number of strings (or sequences) we can build under the constraints of the edge matrix. 

\begin{figure}
    \centering
    \includegraphics[width=0.6\linewidth]{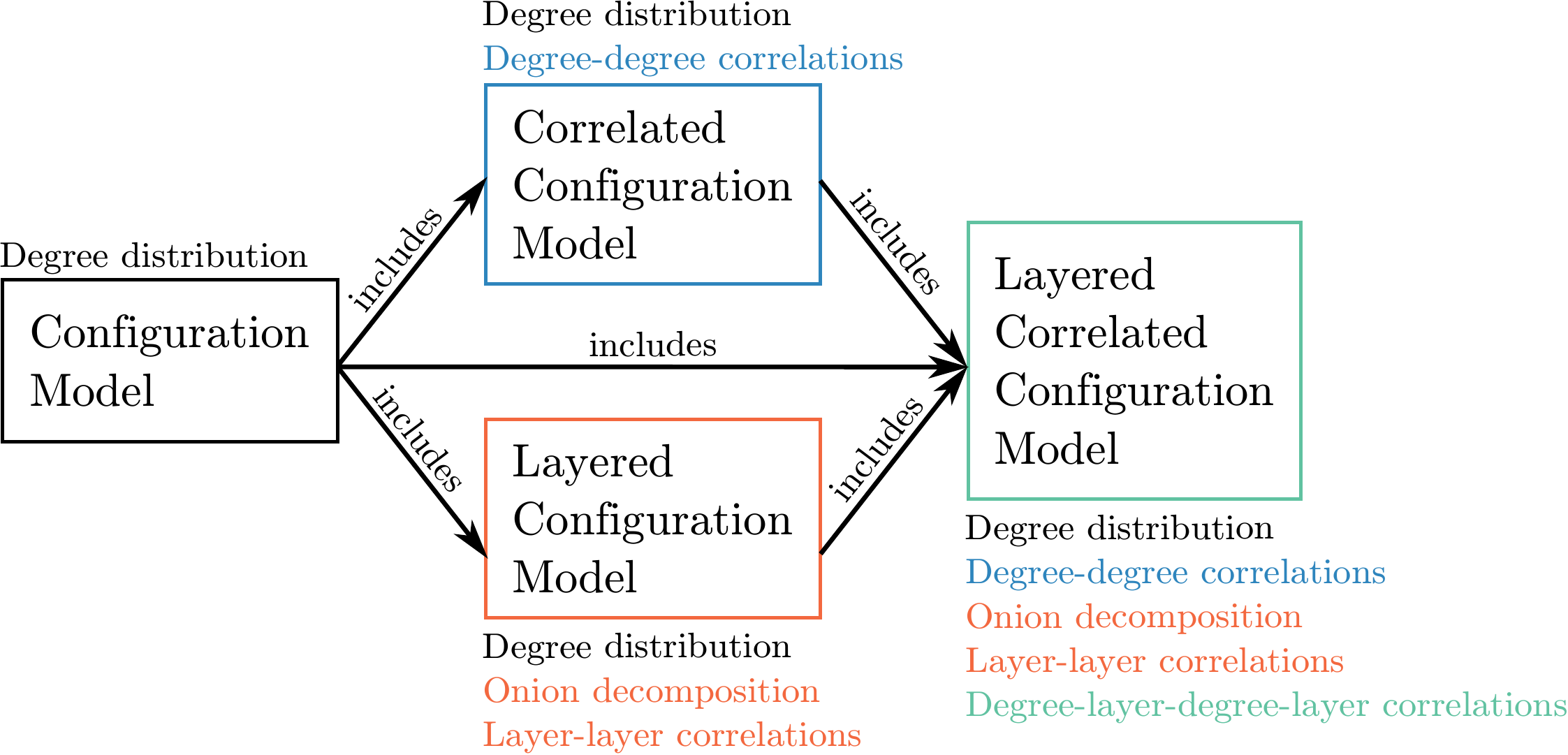}
    \caption{\textbf{Inclusion relationships of our configuration models.} The Configuration Model consists of all simple graphs with a fixed degree sequence and is a superset of all of the other models considered in this study. The Layered and Correlated Configuration Models (LCM and CCM respectively) include additional information of a different nature---layer centrality and degree-degree correlations respectively. By combining the constraints in both the LCM and CCM, the LCCM represents a subset of all previous models.}
    \label{fig:schema}
\end{figure}

Finally, since all models are essentially the CM with added node types, correlations, and constraints, we already know that the ensembles of networks generated by the CCM, LCM, and LCCM are all subsets of the CM ensemble. The precise inclusion relationships of our family of configuration models are represented in Fig.~\ref{fig:schema}.

While we conceptually only consider simple graphs with no self-loops or multiple edges between neighbours, the calculations of the next section ignore these constraints to simplify the combinatorics and obtain rough approximation of ensemble sizes. 
This is a bad assumption in the sense that the data we wish to model always represent simple graphs, but it is generally acceptable in the limit of large networks where the probability of finding self-loops or multiple edges goes to zero due to randomness. 
Further, it is more or less necessary, as no exact calculations are known even for the simple CM \cite{noy2015graph,lucatero2019combinatorial}, although there are asymptotic results \cite{mckay1990asymptotic,mckay1991asymptotic,liebenau2017asymptotic}. 
The upshot is that the same approximation method can be applied to more complex models like the LCM and LCCM, for which no results---even asymptotic ones---are known. 

\subsection{Minimum description length}
To assess the quality of the compression provided by these Configuration Models, we use the Minimum Description Length principle~\cite{grunwald2007minimum,mackay2003information,peixoto2013parsimonious}. This principle favors models that provide the best compression of an observed network, i.e., those that allow a sender to communicate its structure to a receiver with the least amount of information.
For example, the Erd\H{o}s-R\'enyi (ER) model discussed in the introduction can be seen as compressing the structure of a network down to $\mathcal{S}_{\mathrm{ER}}=\log_2 \binom{N(N-1)/2}{E}$ bits, since one can associate each of the  $ \Omega_{\mathrm{ER}}=\binom{N(N-1)/2}{E}$ networks of $E$ edges and $N$ nodes to a unique bitstring of length $\mathcal{S}_{\mathrm{ER}}= \log_2  \Omega_{\mathrm{ER}}$, and thus transmit a network with one such string.
We say that the ER \emph{model} provides the compression because the length $\mathcal{S}_{\mathrm{ER}}$ of the bitstring corresponds to the entropy of the ER model, $\mathcal{S}_{\mathrm{ER}} = -\sum ( 1/ \Omega_{\mathrm{ER}}) \log_2 (1/ \Omega_{\mathrm{ER}}) = \log_2  \Omega_{\mathrm{ER}}$. 
Shannon's source coding theorem shows this is the most efficient compression when networks are truly drawn from the ER model~\cite{cover2006elements}. 

We must also consider the parameters to compute the total cost of transmitting an arbitrary network with the coding scheme implied by a model. 
For example, in the case of the ER model, these parameters are the node and edge counts, which take $\log_2 E$ and $\log_2 N$ bits to send (or a bit more if we choose representations of fixed lengths, e.g., 64-bit integers).~\footnote{In what follows, we use $\log$ without subscript as we always work in units of bits.}
In this particular case, the information costs, in bits, of these numerical parameters can be neglected compared to a model's entropy, and we will thus ignore them here.
In general, however, the parameters of more complex models can amount to significant information costs of the same order as $\mathcal{S}$. 
In general, we thus write the compression provided by a model as
\begin{equation}
     \mathcal{L}_{\mathrm{Model}} = \mathcal{S}_{\mathrm{Model}} + \mathcal{P}_{\mathrm{Model}},
\end{equation}
where $\mathcal{P}_{\mathrm{Model}}$ is the number of bits needed to send model parameters.
This additive form shows that a model that does not leave much to randomness may nonetheless be passed upon if it incurs a large parameter entropy $\mathcal{P}_{\mathrm{Model}}$.

%
%
%
%
%
\section{Calculations}
%

\subsection{Configuration Model}
\begin{figure}
    \centering
    \includegraphics[width=0.8\linewidth]{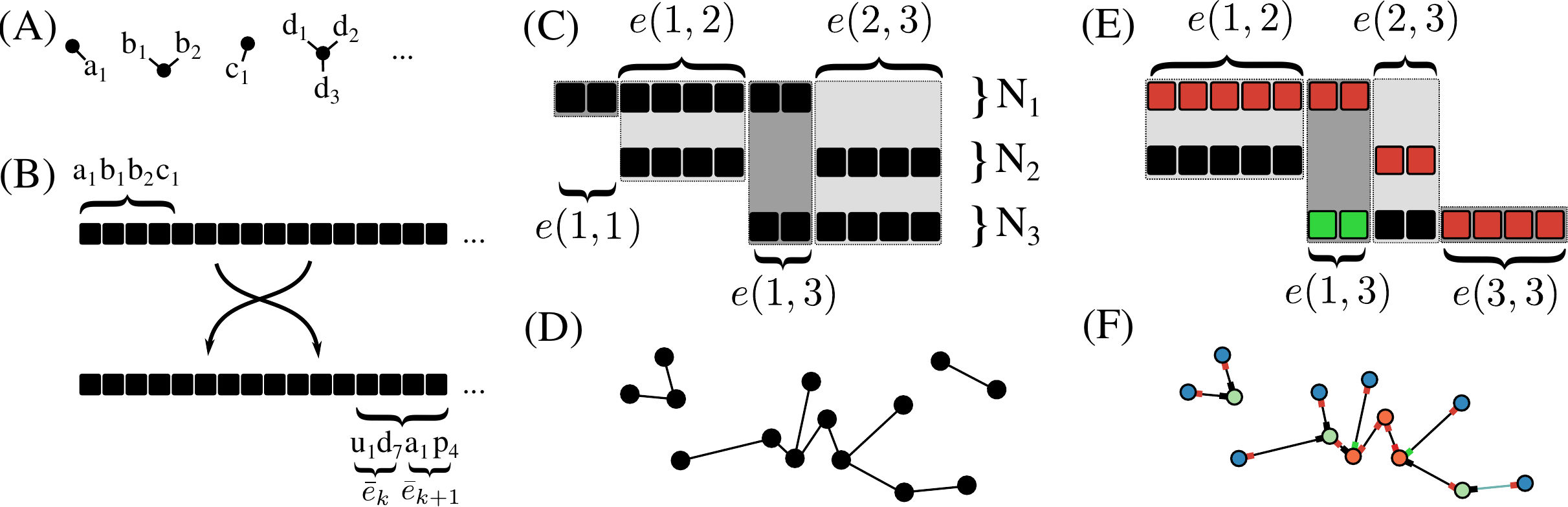}
    \caption{\textbf{Shuffling edges in random networks.}
        (A-B) In the Configuration Model, we can tag stubs according to the node to which they are attached, shown in A. All random orderings of all $2E$ stubs can be turned to networks by connecting adjacent nodes pairwise, shown in B. We must however account for the fact that different orderings can lead to the same network: The order of the stubs of a given node does not matter, the ordering of edges does not matter and the order of stubs within an edge does not matter.
        (C-D) In the Correlated Configuration Model, we now separate the stub list into distinct lists containing the stubs attached to nodes of a given degree, shown in C for the network in D.
        (E-F) In the Onion Network Ensemble of Ref.~\cite{hebert2016multi}, the stub list is separated into different lists for stubs attached to nodes of different layers, and stubs from layer $l$ are also colored according to whether they point to $l'<l-1$ (green), $l'=l-1$ (black) and $l'>l-1$ (red). Panel E shows an example of stub lists for the network in F. The colors of the stubs in E are not related to the colors of the nodes in F, which correspond to the layer where the node is found. The Layered Correlated Configuration Model considered in the text extends this description by distinguishing edges not only by the layer they connect (i.e., $e(l,l')$) but by the joint degree-layer type of nodes they connect (i.e., $e(\{k,l\},\{k',l'\})$.}
    \label{fig:cartoon}
\end{figure}

The CM is the simplest model we consider, as there are no constraints or correlations to take into account other than the degree distribution. The network is specified through $T$ non-empty degree classes containing the sets of $N_k$ nodes of degree $k$. The generative process is equivalent to reshuffling a string of tokens representing individual stubs. Since the degree distribution $p_k = N_k/N$ is given, we know we have $N_k$ nodes with $k$ stubs and therefore $N_k$ nodes contributing $k$ tokens in our string for a total of $\sum kN_k = 2E$ tokens. The number of ways to shuffle the string is given by $(2E)!$. We are, however, overcounting some networks, as we can shuffle pairs of tokens and keep the graph intact. This process is illustrated in Fig.~\ref{fig:cartoon}(A-B).

Accounting for the shuffling of pairs, the permutation of token order within pairs, and the equivalence of stubs from the same node, the ensemble size for the CM is given by
\begin{equation} \label{eq:omega_cm}
    \Omega_{\textrm{CM}} = \frac{(2E)!}{E!2^E\prod _k (k!)^{N_k}} \; .
\end{equation}
The term $E!$ controls the number of ways to reshuffle the edges obtained from a given stub sequence without actually changing the network. Similarly, $2^E$ accounts for the possible ways to reshuffle any two stubs within every edge. Finally, $\prod _k (k!)^{N_k}$ comes from a product over every node $i$ with degree $k_i$, i.e., $\prod_i k_i!$, which corresponds to the number of possible ways to reshuffle all stubs corresponding to the same nodes. Note that we are still considering nodes as labeled, meaning that we do not attempt to control for isomorphic realizations of the models: If two networks are identical when ignoring node labels, they are still counted as separate realizations in this calculation. 

Combining Eq.~\eqref{eq:omega_cm} with $S = \log \Omega$ yields the following micro-canonical entropy for the CM ensemble
\begin{equation} \label{eq:s_cm}
    \mathcal{S}_{\textrm{CM}} = \log \left[(2E)!\right] - \log \left[E!\right] - E \log 2 - \sum _k N_k \log \left[k!\right] \; .
\end{equation}

To obtain the description length $\mathcal{L}_{\textrm{CM}}$ of the CM, we add the information needed to define the model itself~\cite{grunwald2007minimum,mackay2003information}. In this case, the only information required is the degree distribution. The cost of this information can be upper-bounded by its microcanonical entropy \cite{peixoto2013parsimonious}, the logarithm of the number of possible degree sequences given the already known constraints $N$ and $T$. This is the length in bits (or nats, if using the natural logarithm) of the bitstring one needs to send to a receiver to specify a particular degree sequence out of the total number of possible degree sequences with $N$ nodes and $T$ unique degree classes. In this case, one can find an upper bound on the microcanonical entropy. We first specify a degree distribution out of all possible histograms of $N$ integers in the 1 to $T$ interval representing a distribution over non-empty degree classes \cite{peixoto2017nonparametric}, i.e., $\multiset{T}{N} = \binom{N+T-1}{N}$. This gives us a degree distribution but not the individual type of every labeled node. We, therefore, also need to specify one of the $\frac{N!}{\prod_{k}N_k!}$ permutations of node labels that lead to the same labeled type sequence over the chosen histogram.
We finally need to specify which degree should be assigned to each of the $T$ non-empty degree node classes, which is one of $\binom{k_\textrm{max}+1}{T}$ possible assignments, where $k_\textrm{max}$ is the maximum degree in the network. Summing all of these terms allows us to approximate the number of bits needed to parametrize the CM as
\begin{align} \label{eq:p_cm}
    \mathcal{P}_{\textrm{CM}} = \mathcal{S}_{\textrm{CM}} +& \log\left[(T+N-1)!\right] - \log\left[N!\right] - \log\left[(T-1)!\right] \nonumber \\
    &+ \log\left[N!\right] - \sum_k \log\left[N_k!\right] + \log\left[(k_\textrm{max}+1)!\right] - \log\left[T!\right] - \log\left[(k_\textrm{max}+1-T)!\right]\; .
\end{align}
The DL of the classic CM is given by the sum of Eqs. (\ref{eq:s_cm}) and (\ref{eq:p_cm}),
\begin{align} \label{eq:l_cm}
    \mathcal{L}_{\textrm{CM}} = \mathcal{S}_{\textrm{CM}} + \mathcal{P}_{\textrm{CM}} \; .
\end{align}

\subsection{Correlated Configuration Model}
The CCM is an extension of the CM with added degree-degree correlations. To include these correlations, each degree class is a unique node type. The edge matrix $\bm{e}$ then provides the number $e(k,k')$ of edges between nodes of degree $k$ and nodes of degree $k'$.

As illustrated in Fig.~\ref{fig:cartoon}(C-D), we are now shuffling a stub list for each node type, or degree class, meaning that we have at most $k_{\textrm{max}}$ lists of tokens. We then connect nodes both across lists (for instance, the first $e(1,2)$ tokens in the 2nd list are stubs from nodes of degree 2 that will be connected to stubs from nodes of degree 1) and within the lists (diagonal elements $e(k,k)$). There are $\prod (kN_k)!$ ways of shuffling all the lists and accounting for the same permutations as before, giving
\begin{equation} \label{eq:omega_ccm}
    \Omega _{\textrm{CCM}} = \frac{\prod_k \left(kN_k\right)!}{\prod _{k,k'>k} \left[e(k,k')!\right]\prod _k \left[2^{e(k,k)}e(k,k)!(k!)^{N_k}\right]} \; .
\end{equation}
The first product in the denominator accounts for the shuffling of edges between tokens of different lists. The factors in the second product of the denominator account respectively for the permutation of token order within $e(k,k)$ edges, for the shuffling of such edges, and for the equivalence of stubs of the same node. Taking the log, we then find
\begin{equation} \label{eq:s_ccm}
    \mathcal{S} _{\textrm{CCM}} = \sum_k \bigg\lbrace\log \left[(kN_k)!\right] - N_k\log \left[k!\right] - e(k,k) \log 2 - \sum _{k'\geq k} \log \left[e(k,k')!\right] \bigg\rbrace\; .
\end{equation}

The description length $\mathcal{L}_{\textrm{CCM}}$ will be similar to $\mathcal{L}_{\textrm{CM}}$ calculated in Eq.~(\ref{eq:l_cm}). To define the CCM model, we must also communicate the $T\times T$ matrix of edges $e(k,k')$. Following Ref. \cite{jerdee2024improved}, we specify which matrix we are using among all possible matrices with marginals $\lbrace kN_k\rbrace$. The information needed to specify which matrix defines the model is then given by the logarithm of the total number of matrices, yielding
\begin{align}
    \mathcal{P}_{\textrm{CCM}} = \log\left[M(\bm{e})\right] & + \log\left[(T+N-1)!\right] - \log\left[N!\right] - \log\left[(T-1)!\right] \nonumber \\ 
    &  + \log\left[N!\right] - \sum_k \log\left[N_k!\right] +  \log\left[(k_\textrm{max}+1)!\right] - \log\left[T!\right] - \log\left[(k_\textrm{max}+1-T)!\right]\; ,
     \label{eq:p_ccm}
\end{align}
where $M(\bm{e})$ is the number of matrices with the marginals of the network edge matrix $\bm{e}$ as calculated in Ref. \cite{jerdee2024improved}. The remaining terms specify the degree sequence as in our previous calculation for the CM. 
Summing Eqs. (\ref{eq:s_ccm}) and (\ref{eq:p_ccm}), we get the following DL for the CCM,
\begin{align} \label{eq:l_ccm}
    \mathcal{L}_{\textrm{CCM}} = \mathcal{S}_{\textrm{CCM}} + \mathcal{P}_{\textrm{CCM}} \; .
\end{align}

\subsection{Layered Configuration Model}
\begin{figure}
\centering
\includegraphics[width=0.4\linewidth]{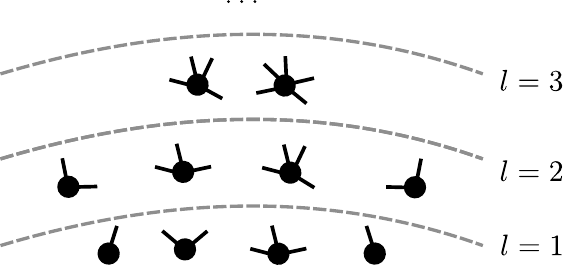}
\caption{\textbf{Layered Configuration Models.} In Layered Configuration Models, nodes are described by their joint layer-degree type. In the models considered here, edges follow constraints that preserve the layer of a node (i.e., its centrality) under the Onion Decomposition \cite{hebert2016multi}. Following these local connection rules provides an edge shuffling mechanism that allows us to produce random networks with a fixed centrality structure based on the concept of $k$-cores and onion layers.}
\label{fig:lcm}
\end{figure}

The LCM is the model used to generate the Onion Network Ensemble in Ref.~\cite{hebert2016multi}. The layers are used to enforce a centrality structure in the random networks; an illustration of the idea is given in Fig.~\ref{fig:lcm}. This centrality structure is based on the Onion Decomposition, a refined version of the classic $k$-core decomposition. In the $k$-core decomposition, we are looking for $k$-cores: the maximal subset of nodes where all nodes have a degree at least $k$ among each other. To find $k$-cores, the algorithm removes all nodes of degree less than $k$ iteratively until all nodes left have a degree at least $k$. The sets of nodes removed at every iteration of the algorithm define the layers of the Onion Decomposition.
A layer $l$ corresponds to a unique coreness $c(l)$ where $c(l)+1$ equals the $k$ used to remove nodes of the $l$-th layer in the $k$-core algorithm. As described in Ref.~\cite{hebert2016multi}, we know that nodes in layer $l$ have at least $c(l)$ degrees leading to nodes in layers $l'\geq l-1$ and at most $c(l)$ degrees leading to nodes in layers $l'' \geq l$. To keep track of these structural constraints stubs are now colored based on which layers they reach. From layer $l$, stubs leading to layers $l'<l-1$ are called green stubs, stubs leading to layer $l'=l-1$ are called black stubs, and stubs leading to layers $l'\geq l$ are called red stubs.
In the LCM, we specify the joint layer-degree distribution $N_{k,l}/N$ for nodes and a coarse-grained edge matrix $\bm{e}$ that specifies the number $e(l,l')$ of edges between two layers. From that information, we can define the fraction of nodes in layer $l$ and the average degree of nodes in layer $l$: $w_l = \sum _k N_{k,l}/N$ and $\langle k \rangle _l =  \sum _k kN_{k,l} / w_lN$ respectively. These quantities allow us to calculate the fraction of stubs leaving layer $l$ that are red or green using
\begin{align}
  p_{r|l} & = \frac{\sum_{l^\prime \geq l} e(l,l')}{w_l c_l / \langle k \rangle} \\
  p_{g|l} & = \frac{\sum_{l' < l-1}  e(l',l)}{w_l [\langle k \rangle_l - c(l)] / \langle k \rangle}
\end{align}
which yields the colored degree distribution for nodes in layer $l$ of degree $k$:
\begin{align}
  P(k_r,k_g,k_b|l,k) = \displaystyle \binom{c_l}{k_r}& \big[ p_{r|l} \big]^{k_r} \big[ 1 - p_{r|l} \big]^{c_l-k_r} \\
   & \times \binom{k-c_l-\delta_{k_r,c_l}\delta_{c_l,c_{l-1}}}{k_g} \big[ p_{g|l} \big]^{k_g} \big[ 1 - p_{g|l} \big]^{k-c_l-k_g-\delta_{k_r,c_l}\delta_{c_l,c_{l-1}}} \delta_{k,k_r+k_b+k_g} \; .
\end{align}
A similar derivation is provided in more detail in Ref.~\cite{allard2018percolation}.

Bringing this back to ensemble size, in Fig.~\ref{fig:cartoon}(E-F), we now not only have $l_{\textrm{max}}$ lists (one for each layer), but tokens now adopt a color-based on which layers $l'$ they can reach; whereas stubs in the CCM follow correlations, but are mostly always free to connect to any other type of nodes. We already know the distributions of red, green, and black stubs per layer, and we know that within a given layer $l$, they must respectively sum to
\begin{equation}
    R(l) = \sum _{l'\geq l} e(l',l) + 2e(l,l) \; , \quad G(l) = \sum _{l'<l-1} e(l',l) \; , \quad B(l) = e(l,l-1) \; .
\end{equation}
We can then shuffle $3$ lists for every $l_{\textrm{max}}$ layers, such that every list contains stubs from a specific layer with a specific color/destination. We find
\begin{equation} \label{eq:omega_lcm}
    \Omega _{\textrm{LCM}} = \prod_l\frac{R(l)!G(l)!B(l)!}{2^{e(l,l)}\prod_{l'\geq l} e(l,l')! \prod _{k} \prod_{k_r,k_g,k_b} \left(k_r!k_g!k_b!\right)^{N_{k_r,k_b,k_g\vert k,l}}} \; ,
\end{equation}
as well as
\begin{align} \label{eq:s_lcm}
    \mathcal{S}_{\textrm{LCM}} = \sum_l \bigg\lbrace\log \left[R(l)!\right] + \log \left[G(l)!\right] + & \log \left[B(l)!\right] - e(l,l) \log 2 - \sum _{l'\geq l} \log \left[e(l,l')!\right] \nonumber \\
    & -\sum _{k,k_r,k_b,k_g} N_{k_r,k_b,k_g \vert k,l}\left[\log k_r! + \log k_g! + \log k_b!\right] \bigg\rbrace\; .
\end{align}
To calculate the cost of the corresponding parametrization of the LCM, we again use a similar logic to calculate the description length of the LCM. Note that the edge matrix is of dimension at most $l_{\textrm{max}}$ and different from the total number of node types $T$ specified by the joint degree-layer types of which there are at most $k_\textrm{max}\ell_\textrm{max}$. The description length $\mathcal{L}_{\textrm{LCM}}$ is therefore upper bounded by
\begin{align} 
    \mathcal{P}_{\textrm{LCM}} = \log\left[M(\bm{e})\right] & + \log\left[(T+N-1)!\right] - \log\left[N!\right] - \log\left[(T-1)!\right] + \log\left[N!\right] - \sum_{k,\ell} \log\left[N_{k,\ell}!\right] \nonumber \\ 
    &  + \log\left[(k_\textrm{max}\ell_\textrm{max}+1)!\right] - \log\left[T!\right] - \log\left[(k_\textrm{max}\ell_\textrm{max}+1-T)!\right]\; ,
    \label{eq:p_lcm}
\end{align}
which could be further refined to account for the fact that not all matrices with the marginals of $\bm{e}$ correspond to graphical LCM edge matrices because of the constraints necessary to preserve layer centrality. 
With this current form, we obtain the following DL for the LCM by summing Eqs. (\ref{eq:s_lcm}) and (\ref{eq:p_lcm}),
\begin{align} \label{eq:l_lcm}
    \mathcal{L}_{\textrm{LCM}} = \mathcal{S}_{\textrm{LCM}} + \mathcal{P}_{\textrm{LCM}} \; .
\end{align}

\subsection{Layered Correlated Configuration Model}
The LCCM is an extension of the previous model to also account for degree-degree correlations~\cite{allard2018percolation}. In the LCCM, we specify the joint degree-layer distribution $N_{k,l}/N$ for nodes and an edge matrix $\bm{e}$ defined over all pairs of joint degree-layer node types to specify the number $e(\{k,l\},\{k',l'\})$ of edges between them. As with the previous model, we can define the fraction of nodes in layer $l$ and the average degree of nodes in layer $l$, respectively $w_l = \sum _k N_{k,l}/N$ and $\langle k \rangle _l =  \sum _k kN_{k,l} / w_lN$. These quantities allow us to calculate the fractions of stubs leaving layer $l$ that are red or green using
\begin{align}
  p_{r|k,l} & = \frac{\sum_{l^\prime \geq l}\sum _{k,k'} e(\{k,l\},\{k',l'\})}{N_{k,l} c(l) / N\langle k \rangle} \\
  p_{g|k,l} & = \frac{\sum_{l' < l-1}  \sum _{k,k'} e(\{k,l\},\{k',l'\})}{N_{k,l} [k - c(l)] / N\langle k \rangle}
\end{align}
which yields the colored degree distribution for nodes of degree $k$ in layer $l$:
\begin{align}
  P(k_r,k_g,k_b|k,l) = \displaystyle \binom{c(l)}{k_r}& \big[ p_{r|l} \big]^{k_r} \big[ 1 - p_{r|l} \big]^{c(l)-k_r} \\
   & \times \binom{k-c(l)-\delta_{k_r,c(l)}\delta_{c(l),c(l-1)}}{k_g} \big[ p_{g|l} \big]^{k_g} \big[ 1 - p_{g|l} \big]^{k-c(l)-k_g-\delta_{k_r,c(l)}\delta_{c(l),c(l-1)}} \delta_{k,k_r+k_b+k_g} \nonumber
\end{align}
The mathematical description of the LCCM is developed in detail in Ref.~\cite{allard2018percolation}. As with the LCM, we now deal with red, green, and black colored tokens representing different types of stubs for every degree-layer type,
\begin{align}
    R(k,l) = \sum _{k',l'\geq  l} e(\{k,l\},\{k',l'\}) & + 2e(\{k,l\},\{k,l\}) \; , \quad G(k,l) = \sum _{k',l'<l-1} e(\{k,l\},\{k',l'\}) \; , \nonumber \\ & B(k,l) = \sum _{k'} e(\{k,l\},\{k',l-1\}) \; .
\end{align}
We can then shuffle up to $3k_{\textrm{max}}l_{\textrm{max}}$ lists that account for the degree and layer of a node type as well as the colors of their stubs. We find
\begin{equation} \label{eq:omega_lccm}
    \Omega _{\textrm{LCCM}} = \prod_{k,l}\frac{R(k,l)!G(k,l)!B(k,l)!}{2^{e(\{k,l\},\{k,l\})}\prod_{k'\geq k,l'\geq l} e(\{k,l\},\{k',l'\})! \prod_{k_r,k_g,k_b} \left(k_r!k_g!k_b!\right)^{N_{k_r,k_b,k_g\vert k,l}}} \; ,
\end{equation}
as well as
\begin{align} \label{eq:s_lccm}
    \mathcal{S}_{\textrm{LCCM}} = \sum_{k,l} \bigg\lbrace\log \left[R(k,l)!\right] + \log \left[G(k,l)!\right] + & \log \left[B(k,l)!\right] - e(\{k,l\},\{k,l\}) \log 2 - \sum _{k'\geq k,l'\geq l} \log \left[e(\{k,l\},\{k',l'\})!\right] \nonumber \\
    & -\sum _{k_r,k_b,k_g} N_{k_r,k_b,k_g \vert k,l}\left[\log k_r! + \log k_g! + \log k_b!\right] \bigg\rbrace\; .
\end{align}

The description length $\mathcal{P}_{\textrm{LCCM}}$ is given by an expression almost identical to the previous one but with $g=T$ as edges and nodes are now distributed over the same joint degree-layer types:
\begin{align} 
\mathcal{P}_{\textrm{LCCM}} = \log\left[M(\bm{e})\right] &+ \log\left[(T+N-1)!\right] - \log\left[N!\right] - \log\left[(T-1)!\right] + \log\left[N!\right] - \sum_{k,\ell} \log\left[N_{k,\ell}!\right] \nonumber \\ 
    &  +   \log\left[(k_\textrm{max}\ell_\textrm{max}+1)!\right] - \log\left[T!\right] - \log\left[(k_\textrm{max}\ell_\textrm{max}+1-T)!\right]\; .
\label{eq:p_lccm}
\end{align}
Note that Eq.~\eqref{eq:l_lccm} is again an upper bound since not all $T\times T$ matrices whose integer entries sum to 2E correspond to graphical LCCM edge matrices because of the constraints necessary to preserve layer centrality. 
Finally, the DL for the LCCM is again obtained by summing Eqs. (\ref{eq:s_lccm}) and (\ref{eq:p_lccm}),
\begin{align} \label{eq:l_lccm}
    \mathcal{L}_{\textrm{LCCM}} = \mathcal{S}_{\textrm{LCCM}} + \mathcal{P}_{\textrm{LCCM}} \; .
\end{align}

Using the DL framework, Eqs. (\ref{eq:l_cm}), (\ref{eq:l_ccm}), (\ref{eq:l_lcm}), and (\ref{eq:l_lccm}) summarize how well these different configuration models can describe network data. One can think of these results as the level of compression of actual network data by normalizing them with the number of bits needed to specify exactly the same network with no model through only its edgelist. While the ensemble size generated by a model informs us about the loss incurred during the compression, this ratio informs us about the compression factor of the model. The lower the compression factor, the more compression the model can achieve.
%
%
%
%
%
\section{Results and discussion}
\subsection{Model comparison on network data}
\begin{table}
\begin{tabular}{ @{}l@{\hspace{0.3cm}} l c@{\hspace{0.8cm}}c c@{\hspace{0.8cm}}c c c c@{} }
Network\hspace{0.2cm} & Domain & Ref. & $N$ & $\langle k \rangle$ & $\ln \Omega_{\textrm{CM}}$ & $\ln \Omega_{\textrm{CCM}}$ & $\ln \Omega_{\textrm{LCM}}$ & $\ln \Omega_{\textrm{LCCM}}$ \\
\hline
\hline

Cayley tree ($z=3, d=6$) & Synthetic & -- & 190 & 2.0 & 765 & 748 & 503 & {\color{monbleu} 503} \\ 
AdoHealth & Social & \cite{Moody2001a} & 2,539 & 8.2 & 62,651 & 61,491 & 49,228 & {\color{monbleu} 35,749} \\ 
arXiv & Social & \cite{Newman2001c} & 30,561 & 8.2 & 983,426 & 958,272 & 856,298 & {\color{monbleu} 723,576} \\ 
Jazz Musicians & Social & \cite{Gleiser2003} & 198 & 27.7 & 6,537 & 4,337 & {\color{monbleu} 4,310} & 6,312 \\ 
Network Scientists & Social & \cite{Newman2006} & 1,461 & 3.8 & 15,266 & 12,847 & {\color{monbleu} 10,224} & 10,232 \\ 
Slashdot & Social & \cite{Kunegis2009} & 82,168 & 12.3 & 3,382,330 & 3,206,620 & 3,221,360 & {\color{monbleu} 3,076,030} \\ 
C. Elegans Genetic & Biological & \cite{DeDomenico2015} & 3,180 & 3.5 & 30,519 & 28,628 & 26,439 & {\color{monbleu} 24,550} \\ 
E. Coli Metabolism & Biological & \cite{Serrano2012} & 1,010 & 6.5 & 14,788 & 12,896 & 11,682 & {\color{monbleu} 11,624} \\ 
Flu Phylogenetics & Biological & \cite{hadfield2018nextstrain} & 4,022 & 2.0 & 28,301 & 27,806 & 22,449 & {\color{monbleu} 22,162} \\ 
Protein Interactions (yeast) & Biological & \cite{Palla2005} & 2,614 & 4.9 & 36,177 & 34,468 & 28,933 & {\color{monbleu} 23,848} \\ 
Plant Pollinators & Ecological & \cite{robertson1977} & 1,500 & 20.3 & 64,884 & 57,434 & 55,049 & {\color{monbleu} 44,498} \\ 
Drosophila Connectome & Connectome & \cite{Takemura2013} & 1,781 & 10.0 & 35,991 & 30,618 & {\color{monbleu} 30,502} & 31,961 \\ 
German Roads & Infrastructure & \cite{Kaiser2004a} & 1,168 & 2.1 & 7,383 & 7,314 & 5,415 & {\color{monbleu} 5,386} \\ 
USA Grid & Infrastructure & \cite{Watts1998} & 4,941 & 2.7 & 47,036 & 46,587 & 39,287 & {\color{monbleu} 38,556} \\ 
World Airports & Infrastructure & \cite{Kunegis2014} & 2,939 & 10.7 & 66,758 & 54,937 & {\color{monbleu} 54,740} & 57,467 \\ 
Gnutella & Technological & \cite{Matei2002} & 36,682 & 4.8 & 726,491 & 714,576 & 670,025 & {\color{monbleu} 615,599} \\ 
Internet AS & Technological & \cite{Karrer2014} & 6,474 & 3.9 & 61,969 & 55,458 & {\color{monbleu} 54,726} & 76,853 \\ 
PGP & Technological & \cite{Boguna2004} & 10,680 & 4.6 & 162,652 & 150,996 & 120,598 & {\color{monbleu} 110,051} \\ 

\hline
\hline
\end{tabular}

\caption{Sizes of the configuration model ensembles for a few empirical networks. Smallest ensembles are highlighted in  {\color{monbleu}blue}.}
\label{table:sizes}
\end{table}

To provide some intuition about the size of random network ensembles (a proxy for how closely they represent network data) and how their description lengths compare to each other, we calculate Eqs.~(\ref{eq:s_cm}-\ref{eq:l_cm}), (\ref{eq:s_ccm}-\ref{eq:l_ccm}), (\ref{eq:s_lcm}-\ref{eq:l_lcm}), and (\ref{eq:s_lccm}-\ref{eq:l_lccm}) on a few representative networks and show the results in Tables \ref{table:sizes} and \ref{table:dl}. A perfect tree (in this case, a Cayley tree network, made using six layers with coordination number three) provides a good example of how the size of a random network ensemble is inflated through isomorphisms. According to our calculations, the LCM and LCCM lead to a total of $2^{502}$ networks on this perfect tree. One can play with the network and edge shuffling rules to conclude that all of these networks are actually isomorphic to each other, meaning that they only swap node labels but preserve the exact structure of the tree.

More generally, we find in Table \ref{table:sizes} that the most constraining model, the LCCM, almost always leads to the smallest ensemble size. And since the CM, LCM, and CCM are all supersets of the LCCM, the cases that show otherwise are due to small data size and the simple graph approximation in our calculations. In particular, because our calculations rely on the probabilities of self-loops and multiple edges going to zero, we expect over-counting of invalid realizations when there are very few possible pairings of nodes within certain node classes. This over-counting is much more likely in the most detailed model, the LCCM. Ignoring self-loops and multiple edges was necessary to have a flexible calculation methods able to tackle intricate models, but is an approximation that could be improved in future work.

More importantly, despite its small ensemble size, the LCCM does not provide a great compression of network data, as shown in Table \ref{table:dl} because the underlying model is fairly complicated to specify. Interestingly, among the three other simpler models, we find that they all provide the best description on some networks. 
From these few networks, it appears that the CCM is preferred when compressing biological and ecological networks, while the CM and LCM do best in the rest of the networks.
We can hypothesize that since the LCM constrains its description of a centrality structure based on a treelike decomposition of networks, it should perform best on very sparse networks whose structure is more likely to be treelike.

\begin{figure}[!]
\centering 
\includegraphics[width=0.7\linewidth]{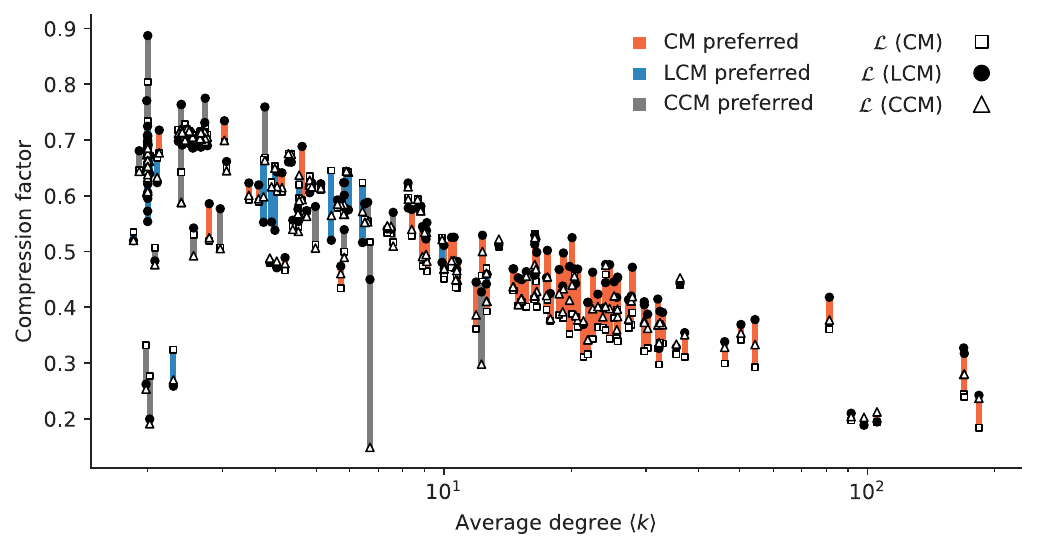}
\caption{\textbf{Compression of over 200 network datasets with varying average degree $\bm{\langle k \rangle}$.} 
The compression factor is defined as the ratio of the description length of the network, assuming a model, to its description length using an edge list, i.e., $2E \log (N)$ bits. 
Smaller values mean the model compresses more.
Every network dataset is represented by a connected triplet: a black marker indicating the compression level $\mathcal{L}_{CM}$ of the Configuration Model, a triangle indicating the level $\mathcal{L}_{CCM}$ of the Correlated Configuration Model, and a white marker indicating the compression $\mathcal{L}_{LCM}$ attained by the Layered Configuration Model.
The difference between the three is shown in blue if the best-performing model is the LCM, in orange if it is the CM, and in grey when it is the CCM.
The LCM offers a more compact description for most treelike networks with a small average degree $\langle k \rangle$, while the simpler configuration model is generally more compressive for denser networks.
}
\label{fig:compression}
\end{figure}

To test the generality of this conclusion, we calculate the description lengths of the CM, CCM, LCM, and LCCM on over 200 network datasets \cite{allard2018percolation}. Our results are summarized in Fig.~\ref{fig:compression}. We find that the CM provides the best compression for $46\%$ of our network data sets, in comparison with $34\%$ best compressed by the LCM, and the remaining $20\%$ being best compressed by the CCM.
However, we find that the CM performs best on only $20\%$ of networks with an average degree below 10. More complicated models are preferred on these sparse networks, with the LCM offering the best performance on $50\%$ of these networks and the CCM on the remaining $30\%$. In general, we can say that the classic CM performs best on dense networks, while the LCM performs best on sparse networks. Meanwhile, the LCCM requires constraints that are too costly to specify relative to the reduction they achieve in the network ensemble size, resulting in the LCCM never providing the best compression for the network datasets studied.

\begin{table}

\begin{tabular}{ @{}l@{\hspace{0.3cm}} l c@{\hspace{0.8cm}}c c@{\hspace{0.8cm}}c c c c@{} }
Network & Domain & Ref. & $N$ & $\langle k \rangle$ & $\mathcal{L}_{\textrm{CM}}$ & $\mathcal{L}_{\textrm{CCM}}$ & $\mathcal{L}_{\textrm{LCM}}$ & $\mathcal{L}_{\textrm{LCCM}}$ \\
\hline
\hline

Cayley tree ($z=3, d=6$)  & Synthetic & -- & 190 & 2.0 & 1,292 & 1,273 & {\color{monbleu} 1,183} & 1,221 \\ 
AdoHealth & Social & \cite{Moody2001a} & 2,539 & 8.2 & {\color{monbleu} 100,782} & 100,991 & 102,106 & 147,587 \\ 
arXiv & Social & \cite{Newman2001c} & 30,561 & 8.2 & 1,551,680 & 1,545,650 & {\color{monbleu} 1,504,020} & 2,181,670 \\ 
Jazz Musicians & Social & \cite{Gleiser2003} & 198 & 27.7 & {\color{monbleu} 10,693} & 11,922 & 12,175 & 28,045 \\ 
Network Scientists & Social & \cite{Newman2006} & 1,461 & 3.8 & 26,573 & 23,889 & {\color{monbleu} 22,068} & 28,310 \\ 
Slashdot & Social & \cite{Kunegis2009} & 82,168 & 12.3 & 5,216,150 & {\color{monbleu} 3,400,820} & 4,879,990 & 8,908,640 \\ 
C. Elegans Genetic & Biological & \cite{DeDomenico2015} & 3,180 & 3.5 & {\color{monbleu} 52,658} & 53,387 & 55,360 & 80,818 \\ 
E. Coli Metabolism & Biological & \cite{Serrano2012} & 1,010 & 6.5 & 25,322 & {\color{monbleu} 25,103} & 26,630 & 40,895 \\ 
Flu Phylogenetics & Biological & \cite{hadfield2018nextstrain} & 4,022 & 2.0 & 46,092 & {\color{monbleu} 45,478} & 45,866 & 50,651 \\ 
Protein Interactions (yeast) & Biological & \cite{Palla2005} & 2,614 & 4.9 & {\color{monbleu} 61,549} & 61,847 & 62,540 & 94,168 \\ 
Plant Pollinators & Ecological & \cite{robertson1977} & 1,500 & 20.3 & 102,239 & {\color{monbleu} 101,760} & 104,598 & 187,672 \\ 
Drosophila Connectome & Connectome & \cite{Takemura2013} & 1,781 & 10.0 & {\color{monbleu} 60,203} & 62,463 & 68,152 & 120,477 \\ 
German Roads & Infrastructure & \cite{Kaiser2004a} & 1,168 & 2.1 & {\color{monbleu} 11,865} & 11,880 & 12,594 & 13,822 \\ 
USA Grid & Infrastructure & \cite{Watts1998} & 4,941 & 2.7 & 80,077 & 79,982 & {\color{monbleu} 77,948} & 90,731 \\ 
World Airports & Infrastructure & \cite{Kunegis2014} & 2,939 & 10.7 & {\color{monbleu} 109,134} & 112,463 & 120,821 & 224,787 \\ 
Gnutella & Technological & \cite{Matei2002} & 36,682 & 4.8 & 1,160,690 & 1,149,450 & {\color{monbleu} 1,142,220} & 1,382,090 \\ 
Internet AS & Technological & \cite{Karrer2014} & 6,474 & 3.9 & {\color{monbleu} 105,787} & 107,730 & 105,960 & 183,189 \\ 
PGP & Technological & \cite{Boguna2004} & 10,680 & 4.6 & 268,909 & 266,742 & {\color{monbleu} 253,635} & 388,913 \\ 

\hline
\hline
\end{tabular}

\caption{Description length provided by the various configuration models, for a few empirical networks. The smallest description lengths are highlighted in {\color{monbleu}blue}.}
\label{table:dl}
\end{table}

%
%
%
%
%
\subsection{Discussion}
%

The classic Configuration Model is used throughout network science as a simple way of capturing one key distinguishing feature of complex network data, namely their degree heterogeneity. Applications of random networks tend to build on the Configuration Model in one way or another to represent the structure of complex networks at a low cost. Dynamical models are often based on heterogeneous mean-field approximations \cite{pastor2001epidemic}, which assign a dynamical variable to each degree class in the Configuration Models. There is a conceptual simplicity to these models, as the degree distribution is easy to measure and random networks are easy to generate under this constraint---as opposed to, say, inference frameworks like the degree-corrected Stochastic Block Model \cite{peixoto2013parsimonious} that infer block partitions by minimizing the description length over many possible partitions of a network~\cite{peixoto2017nonparametric}. But is the simpler Configuration Model actually a parsimonious description of network data? Are there similar alternatives that better represent complex networks?

We answered this question by analytically approximating the description lengths of the Configuration Model and three of its variants that are further constrained for degree correlations and layered structure. Surprisingly, we found that the Layered Configuration Model \cite{hebert2016multi} offered a great trade-off in model complexity when describing sparse networks. The Layered Configuration Model relies on a simple description of network structure based on the Onion Decomposition, classifying nodes based on their degree and centrality layer but requiring an intricate connection scheme to generate complex random networks from that description.

Our results motivate further study of layered network models as they offer a powerful way of capturing the impact of network centrality on dynamical processes. In that vein, Ref.~\cite{allard2018percolation} showed that adapting a probability-generating function formalism for the LCCM greatly improved predictions of bond percolation thresholds.  Also, Ref.~\cite{hebert2023hierarchical} recently illustrated how a heterogeneous mean-field approach generalized to the LCM allowed for a simple yet effective way to include complex hierarchies governing how people interact within an organization.  These results suggest that other modeling approaches could improve accuracy by including information about the onion decomposition. For instance, Stochastic Block Models could be degree-corrected and layer-corrected to reproduce centrality structure. One could then imagine inferring mesoscopic groupings through block structure while constraining for macroscopic hierarchical structure through onion layers.

Finally, it appears that, by using a complex connection scheme, Layered Configuration Models effectively constrain the space of random networks at a low information cost, allowing for an effective representation of complex network data. 
Hence, simpler, more efficient layered models should be investigated, as the LCM used in this work might not be the most parsimonious layered model possible. More precise calculations of the ensemble produced by these models should also be explored to offer a more accurate model comparison that avoids counting isomorphic realizations. In conclusion, from dynamical systems to inference frameworks and network analysis, we think layered random networks present unique opportunities to reconsider how we summarize and study complex networks. More broadly, our results provide an inspiring example that creative network representations can still outperform simple established models.

%
\section*{Acknowledgements}

The authors thank Joshua Grochow for comments on the project. L.H.-D. and A.D. acknowledge support from the National Science Foundation Grant No. DMS-1829826. 
A.K. was supported by the HKU-100 Start-Up Grant and the Hong Kong Research Grants Council under Grant No. ECS--27302523.
A.A. acknowledges financial support from the Sentinelle Nord initiative of the Canada First Research Excellence Fund and from the Natural Sciences and Engineering Research Council of Canada (project 2019-05183). 

%
%
%
%

%
%
\end{document}